\documentstyle[12pt]{article}

\begin{document}

\title{Relaxation dominated cosmological expansion}

\author{Luis P. Chimento and Alejandro S. Jakubi      \\
{\it Departamento de F\'{\i}sica,  }\\
{\it  Facultad de Ciencias Exactas y Naturales, }\\
{\it Universidad de Buenos Aires }\\
{\it  Ciudad  Universitaria,  Pabell\'{o}n  I, }\\
{\it 1428 Buenos Aires, Argentina.}}

\maketitle

\begin{abstract}

The behavior near the singularity of an isotropic, homogeneous cosmological
model with a viscous fluid source is investigated. This turns out to be
a relaxation dominated regime.  Full extended irreversible thermodynamics
is used, and comparison with results of the truncated theory is made. New
singular behaviors are found and it is shown that a relaxation dominated
inflationary epoch may exist for fluids with small heat capacity.

\end{abstract}

\vskip 3cm


\newpage

There are a many processes capable of producing important dissipative stresses
in the early universe. This includes interactions between
matter and radiation \cite{Wei71}, quarks and gluons \cite{Tho}, different components of dark
matter \cite{Pav93}, and those mediated by massive particles. It happens also due to the
decay of massive superstrings modes \cite{Myu}, gravitational particle
production \cite{Hu} \cite{Bar88} and phase
transitions. Phenomenologically, these processes may be modeled in terms of a
classical bulk viscosity. The dynamics of the viscous pressure term is ruled
by a transport equation, and we use the expression given by the Extended
Irreversible Thermodynamics theory \cite{Maar} \cite{Zak93b}.

\begin{equation} \label{1}
\sigma+\tau\dot\sigma=-3\zeta H-\frac{\epsilon}{2}\tau\sigma
\left(3H+\frac{\dot\tau}{\tau}-\frac{\dot\zeta}{\zeta}-\frac{\dot T}{T}\right)
\end{equation}

\noindent where we are assuming a spatially flat Robertson-Walker metric,
$\sigma$ is the bulk viscous pressure, $H=\dot a/a$ the Hubble variable,
$\tau$ is the relaxation time, $\zeta$ is the bulk viscous coefficient, $T$ is
the temperature, and $\epsilon=0$ corresponds to the truncated theory while
$\epsilon=1$ corresponds to the full theory. Equation (\ref{1}) yields a
causal and stable behavior due to the memory mechanism provided by the
relaxation time. We assume that $\zeta=\alpha\rho^m$, $\tau=\zeta/\rho$ and
$T=\kappa\rho^r$, where $\alpha>0$, $m$, $\kappa>0$ and $r>0$ are constants
\cite{Rom}.

We will examine in this Letter the evolution of the dissipative universe near
a singularity, which is a relaxation
dominated era. In effect, the Einstein equations are

\begin{equation} \label{2}
H^{2} = {\frac{1}{3}} \rho
\end{equation}

\begin{equation} \label{3}
\dot H+ 3H^{2} = {\frac{1}{2}} (\rho - p -\sigma )
\end{equation}

\noindent where $\rho $ is the energy density, $p$ is the equilibrium pressure
and we take the equation of state $p = ( \gamma -1 ) \rho$ with a constant
adiabatic index $0\le\gamma\le 2$. For a singularity in $t_0$, we will
consider two types of leading behaviors when $t\to t_0$, $H\sim 1/\Delta t^n$,
where $n>0$ is a constant, and $H\sim 1/(\Delta t\ln\Delta t)$. For $H\sim
1/\Delta t^n$, $\sigma\propto H^2\sim 1/(\Delta t)^{2n}$ for $n\geq 1$, and
$\sigma\propto \dot H\sim 1/(\Delta t)^{n+1}$ for $n\leq 1$. So,
$|\sigma/(\zeta H)|\sim \Delta t^{n(2m-1)}$ for $n\geq 1$, and $|\sigma/(\zeta
H)|\sim \Delta t^{2nm-1}$ for $n\leq 1$. Also
$\dot\tau/\tau\sim\dot\zeta/\zeta\sim\dot T/T\sim 1/\Delta t$. Thus we find
that the first term of the left hand side of (\ref{1}) becomes much smaller
than the rest of the terms provided that $m>1/2$ and $\tau\gg |\Delta t|$.
Similar considerations are valid for the second type of leading behavior.
Then, using (\ref{1}), (\ref{2}) and (\ref{3}), and neglecting the viscous
term $\sigma$, we obtain

\begin{equation} \label{4}
\ddot H-\epsilon(1+r)\frac{\dot H^2}{H}+
3\left\{\gamma+\frac{\epsilon}{2}\left[1-\left(1+r\right)\gamma
\right]\right\} H\dot H+\frac{9}{4}\left(\epsilon\gamma-2\right) H^3=0
\end{equation}

\noindent   We investigate the solutions of this equation with one of these two 
types of behaviour  near the
singularity, as the solutions of the non-approximated equation will have 
the same leading behavior for
$\Delta t\to 0$. The study of these solutions for large time will be made
elsewhere  (see also \cite{Maar}). As the relaxation regime has
already  been considered in the truncated theory \cite{Zak93a}, we will
concentrate on the case $\epsilon=1$. We also note that an equation with the
same structure arises when considering the semiclassical Einstein equation
with vacuum polarization terms near the singularity \cite{And}.

To verify the consistency of our approximation, we check first whether
equation (\ref{4}) has solutions of the form  $a=a_0\Delta t^\nu$. In effect
there is a region of the parameter plane $(\gamma,r)$, such
that two families of solutions of this form exist, with $\nu$ given by

\begin{equation} \label{5}
\nu_{\pm}=\frac{1}{3(2-\gamma)}\left\{-\left[1+(1-r)\gamma\right]\pm
\left[\left(1+(1-r)\gamma\right)^2-4(2-\gamma)(r-1)\right]^{1/2}\right\}
\end{equation}

\noindent This region is determined by  the requirement $D>0$, where
$D$ is the discriminant in (\ref{5}), and this condition can be satisfied
for $\gamma<2$. On the curve $D=0$ there is only one family of solutions,
and for $\gamma=2$ there is also one family of solutions with
$\nu_2=2(1-r)/(3(3-2r))$.

To investigate further the two-parameter families of solutions of (\ref{4}), it is
convenient to make the change of variable $H=y^{-1/r}$, which turns this
equation into

\begin{equation} \label{6}
\ddot y+\frac{3}{2}\left[1+(1-r)\gamma\right] y^{-\frac{1}{r}}\dot
y+\frac{9}{4}(2-\gamma)ry^{1-\frac{2}{r}}=0
\end{equation}

\noindent In  the case that   $1+(1-r)\gamma\neq 0$, $r\neq
1$ we solve (\ref{6}) applying a generalization of the transformation
used in \cite{Chi93}

$$
z=\frac{r}{r-1}y^{\frac{r-1}{r}},\qquad
d\eta=\frac{3}{2}\left[1+(1-r)\gamma\right] y^{-\frac{1}{r}}dt
$$

\noindent
which linearizes (\ref{6})

\begin{equation} \label{7}
\frac{d^2z}{d\eta^2}+\frac{dz}{d\eta}+\beta z=0
\end{equation}

\noindent
where $\beta=(2-\gamma)(r-1)/\left[1+(1-r)\gamma\right]^2$. Thus, we obtain
the general solution of (\ref{4}) in parametric form

$$
H(\eta)=\left[\frac{r-1}{r}z(\eta)\right]^{-\frac{1}{r-1}}
$$

\begin{equation} \label{8}
\Delta t(\eta)=\frac{2}{3\left[1+(1-r)\gamma\right]}\int d\eta H(\eta)^{-1}
\end{equation}

This parametric form is very useful to obtain the singular behavior in the
region $D\leq 0$, where no explicit solution is available. In this region
$r>1$, so that $H$ diverges when $z\to 0$. Moreover, as this region
corresponds to the oscillating or critically damped regime of the oscillator
described by (\ref{7}), when $\gamma<2$ $z(\eta)\propto \Delta\eta$ for
$\Delta\eta\to 0$. Then we find $H\simeq C/\Delta t^{\frac{1}{r}}$ or

\begin{equation} \label{9}
a(t)\simeq a_0\exp\left(K\Delta t^{\frac{r-1}{r}}\right),\qquad \Delta t\to 0
\end{equation}

\noindent
where $a_0$ and $K$ are arbitrary integration constants. This behavior
exhibits a singularity
at a finite value of the scale factor.

In the case that $1+(1-r)\gamma= 0$ and $\gamma<2$, equation (\ref{6})
 describes the motion of a point particle in a potential $V(y)\propto
 y^{2(r-1)/r}$ and reduces to quadratures

\begin{equation} \label{9.1}
\Delta t=-\frac{r}{\sqrt{2}}\int
dH\frac{H^{-r-1}}{\left[E-GH^{2(r-1)}\right]^{1/2}}
\end{equation}

\noindent
where $E$ is an integration constant and
$G=9(2-\gamma)(1+\gamma)^2/(8\gamma)$. This solution may be expressed
in terms of the hypergeometric function, and we find that the singular
behavior is also (\ref{9}), where $r=(1+\gamma)/\gamma$.

On the other hand, for $D>0$, explicit two-parameter families of
solutions of (\ref{6}) exist provided that the constrain \cite{Chi95}

\begin{equation} \label{10}
\frac{(2-\gamma)r}{\left[1+(1-r)\gamma\right]^2}=
\left(2-\frac{1}{r}\right)^{-2}
\end{equation}

\noindent is satisfied. They yield a representative sample of two-parameter
families of solutions in
this region, and have the form

\begin{equation} \label{11}
a(t)=a_0\left|\left|\Delta t\right|^{\frac{r-1}{r}}+K\right|^{\nu_\gamma},
\qquad r\neq 1
\end{equation}

\begin{equation} \label{12}
a(t)=a_0\left|\ln\left|\Delta t\right|+K\right|^{-\frac{2}{3}},\qquad r=1
\end{equation}

\noindent
where $\nu_\gamma=2(1-2r)/[3(1+(1-r)\gamma)]$. We find in this way four kinds
of singular behaviors, as shown in the following table:

$$
\begin{tabular}{|c|c|c|c|}
\hline
$a(t)$&$\lim a(t)$&Parameter range\\
\hline
$\Delta t^{\nu_+}$&$0$&$0<r<1$ and $r>3/2$\\
\hline
$\Delta t^{\nu_-}$&$\infty$&$0<r<1$ and $1<r<3/2$\\
\hline
$\left|\ln\Delta t\right|^{\nu_-}$&$0$&$r=1$\\
\hline
$\exp(K\Delta t^{\frac{r-1}{r}})$&$a_0$&$r>1$\\
\hline
\end{tabular}
$$

Finally, for $\gamma=2$, equation (\ref{6}) also reduces easily to
quadratures

\begin{equation} \label{}
\Delta t=r\int dH \frac{H^{-r-1}}{SH^{1-r}-E}
\end{equation}

\noindent where $S=3r(3-2r)/(2r-2)$, and this solution be expressed in terms
of the hypergeometric function. We find that their singular behavior is
$\Delta t^{\nu_2}$ for $0<r<1$, $\Delta t^{-2/3}$ for $r=1$, and (\ref{9}) for
$r>1$.

We have investigated the singular behavior of the full causal model for
$m>1/2$. Similarly to the truncated causal model, we find that it is dominated
by the relaxation terms of the transport equation. However we have found that
a greater variety of singular behaviors appear in the full model, and it is
the dependence of the energy density on the temperature which determines the
kind of singular behavior of two-parameter families of solutions.

Big-bang and explosive singularities already appeared  in the truncated
theory. Following the same steps as before and assuming $H=\bar\nu/\Delta t$
we find
$\bar\nu_{\pm}=(1/3)\left[\gamma\pm\left(\gamma^2+4\right)^{1/2}\right]$ (cf.
\cite{Zak93a}). We note that $0<\nu_+<\bar\nu_+<2/(3\gamma)$ for $0<r<1$,
while $\nu_+\ge 4/3>\bar\nu_+$ and $\nu_+>2/(3\gamma)$ for $r>3/2$. Thus we
can state that for a fluid with large heat capacity, the rate of expansion  of
the scale factor after a Big-bang singularity is smaller than calculated in
the truncated theory which is turn smaller than the perfect fluid rate. In
this case, there are particle horizons.
However, when the heat capacity is small, the rate of expansion turns out much
larger for a viscous fluid than a perfect fluid. Moreover, it turns out to be
a relaxation dominated inflationary regime.

On the other hand,
$\nu_-<\bar\nu_-<0$
for $0<r<r_c$ and $\bar\nu_-<\nu_-<0$ for $r_c<r<3/2$, where $r_c$ depends on
$\gamma$.
Thus, explosive singularities may be stronger or milder depending on the heat
capacity being large or small.
A finite scale factor singularity arises for $r>1$, and a logarithmic
explosive singularity occurs for $r=1$. Neither of them appear in the
truncated model, though finite scale singularities has been found in noncausal
models for $m>1/2$ \cite{Sus} \cite{San}.

\end{document}